\documentclass{PoS}
\usepackage{dsfont}

\title{Massless Schwinger model with a 4-fermi interaction at topological angle $\theta = \pi$
}

\ShortTitle{Schwinger model at $\theta = \pi$}

\author{Dominic Hirtler$^{\,a}$\footnote{Speaker.} $\! \!$ , Christof Gattringer$^{\,b}$\footnote{Currently on leave of absence from University of Graz, 8010 Graz, Austria.} 
\vskip4mm

$^{a}$ University of Graz, 8010 Graz, Austria\footnote{Member of NAWI Graz.}

\vskip2mm

$^b$ FWF - Austrian Science Fund, 1090 Vienna, Austria 

\vskip5mm
        
\email{dominic.hirtler@edu.uni-graz.at}

\email{christof.gattringer@fwf.ac.at}   }

\abstract{We study the massless Schwinger model with an additional 4-fermi interaction and a topological term. For topological angle $\theta = \pi$
charge conjugation symmetry is implemented in a non-trivial way and we study the possibility of its spontaneous breaking. For the lattice discretization 
we use staggered fermions and the Villain action for the gauge fields, where the topological term is an integer and charge conjugation at  $\theta = \pi$ is an 
exact symmetry. The complex action problem is overcome by a suitable worldline/worldsheet representation. We find that as a function of the 4-fermi coupling 
the system shows a critical point separating a weak coupling phase where charge conjugation symmetry is intact from a strong coupling phase with 
spontaneously broken charge conjugation symmetry.}

\FullConference{The 39th International Symposium on Lattice Field Theory,\\ 
8th-13th August, 2022,\\
Rheinische Friedrich-Wilhelms-Universität Bonn, Bonn, Germany}

\begin{document}

\section{Introductory comments}

\noindent
Topological terms are an interesting ingredient in quantum field theories, as they may alter the symmetry content of a theory in a non-local way. 
This non-local character implies that a non-perturbative regularization has to be used. An interesting non-perturbative regularization is the lattice, 
which, however, poses two challenges: a suitable discretization of the topological charge, and a way to overcome the complex action problem that 
is caused by the topological term. 

In recent work \cite{Sulejmanpasic:2019ytl} it was shown that a generalization of the Villain action \cite{Villain:1974ir} gives rise to an integer-valued 
definition of the topological charge in terms of the Villain variables. As a consequence, charge conjugation (C) symmetry at 
$\theta = \pi$ is implemented exactly 
and its spontaneous breaking was studied in gauge Higgs models \cite{Gattringer:2018dlw,Goschl:2018uma}, while a study of C symmetry breaking
in the same model but with Wilson gauge action and a field theoretic non-integer definition of the topological charge led to less conclusive results
\cite{Kloiber:2014dfa,Gattringer:2015baa}. We remark that also the Atiyah Singer index theorem \cite{Atiyah:1968mp} is relevant 
for the physics of the fermionic system studied here, which emerges in the continuum limit of the Villain formulation \cite{Gattringer:2019yof}. 
Generalized Villain formulations were also used to map the 
gauged XY model in the strong coupling limit at $\theta = \pi$ to the Ising model \cite{Sulejmanpasic:2020ubo}, to construct a lattice discretization for 
fracton theories \cite{Gorantla:2021svj} and to explore non-invertible duality defects \cite{Choi:2021kmx}.

As was already mentioned, the topological term also generates a complex action problem, which, however, may be solved by switching to a 
worldline/worldsheet representation \cite{Gattringer:2018dlw,Goschl:2018uma} that can be simulated efficiently \cite{Prokofev:2001ddj,Mercado:2013yta}. 
For fermions there is also a potential sign problem coming from 
the Grassmann nature of the fermionic variables and the $\gamma$-algebra. For the case of massless staggered fermions in 2d, i.e., the discretization
we use here, the sign problem is known to be absent \cite{Gattringer:2015nea,Goschl:2017kml}, and also the quartic fermion self interaction does
not alter this result.

The ingredients outlined in the last two paragraphs, i.e., the integer-valued Villain-based definition of the topological charge, the worldline/worldsheet 
representation for overcoming the complex action problem and the absence of a fermionic sign problem for massless staggered fermions in 2d 
allow one for the first time to study the spontaneous breaking of charge conjugation in a fermionic system: The 2d massless Schwinger model with a quartic
self interaction and a topological term at $\theta = \pi$. Charge conjugation appears as a $\mathds{Z}_2$ symmetry and we explore whether it can be broken 
spontaneously as a function of the quartic coupling parameter $J$. 

Using Monte Carlo simulations of the system in its worldline/worldsheet representation we study various bulk quantities, in particular the C symmetry 
breaking topological charge density $\langle q \rangle$ and the corresponding susceptibility. For weak coupling $J$ the symmetry remains unbroken, while 
at strong coupling we observe breaking of C symmetry. We find strong evidence for a critical point near $J \sim 0.9$, which seems compatible with the 2d 
Ising universality class as expected. Varying $\theta$ in the strong coupling phase we observe a first order jump in the order parameter 
$\langle q \rangle$ when crossing $\theta = \pi$, which is a further indication that the system implements the 2d Ising phenomenology.

\section{The Schwinger Model and its worldline/worldsheet representation}

\noindent
The partition sum of our model is given by,
\begin{equation}
Z \, = \; \int \!\!D[A] \, \int \!\!  D\big[\, \overline{\psi}, \psi\big] \, B_{\beta, \theta}[A] \, e^{\, - \, S_F\big[ \, \overline{\psi}, \psi, A \big] }\; ,
\label{eq:Zdef}
\end{equation}
where the path integral measures are the usual product measures over the link-based gauge fields $A_{x,\mu} \in [-\pi, \pi)$ and the site based
Grassmann valued fermion fields $\psi_x$ and $\overline{\psi}_x$,
\begin{equation}
\int \!\!D[A] \; = \; \prod_{x,\mu} \int_{-\pi}^\pi \!\! \frac{dA_{x,\mu}}{2\pi}
\quad , \qquad
\int \!\!  D\big[\, \overline{\psi}, \psi\big] \; = \; \prod_x \int \!\! d \psi_x \, d \overline{\psi}_x \; .
\label{eq:measure}
\end{equation}
The degrees of freedom live on a 2-dimensional lattice with periodic boundary conditions for all fields, with the exception of the temporal boundary
conditions of the fermions which are chosen anti-periodic. The gauge field dynamics is described by the Villain Boltzmann factor \cite{Villain:1974ir}, which,  
following \cite{Sulejmanpasic:2019ytl} is augmented with a topological term (here $(dA)_x = A_{x+\hat1,2} - A_{x,2} - A_{x+\hat2,1} + A_{x,1}$),
\begin{equation}
B_{\beta, \theta}[A] \; = \; \sum_{\{n\}} \, e^{ \, - \, \frac{\beta}{2} \sum_x \big( (dA)_x \, + \, 2 \pi n_x \big)^2 \; - \, i \theta \, Q} 
\;, \qquad Q \; = \; \sum_x n_x
\;\;, \qquad \sum_{\{n\}}  \; = \; \prod_{x} \sum_{n_x \in \mathds{Z}} \; .
\label{eq:Bfactor}
\end{equation}
The Villain variables $n_x \in \mathds{Z}$ are integers assigned to the plaquettes and summed. The topological charge $Q = \sum_x n_x$ is simply 
given by the sum over all Villain variables \cite{Sulejmanpasic:2019ytl} and thus is an integer, a fact that is  
essential for the exact implementation of the charge conjugation symmetry
at topological angle $\theta = \pi$. For the fermions we use the staggered action,
\begin{equation}
S_F \; =  \; \frac{1}{2} \sum_{x,\mu} \gamma_{x,\mu} \left[ \overline{\psi}_x \, e^{\, i \, A_{x,\mu}} \, \psi_{x+\hat \mu} \, - \, 
\overline{\psi}_x \, e^{\, -i \, A_{x-\hat \mu}} \, \psi_{x-\hat \mu} \right]
\; - \; \frac{J}{4} \sum_{x,\mu} \overline{\psi}_x \psi_x 
\overline{\psi}_{x+\hat \mu} \psi_{x+\hat \mu} \; ,
\end{equation}
where the staggered sign factors are given by $\gamma_{x,1} = 1, \gamma_{x,2} = (-1)^{x_1}$. We consider the massless case, but include a quartic
term, such that we can use the coupling $J$ as the control parameter for exploring the possible breaking of charge conjugation symmetry. Note that 
the quartic term is not invariant under chiral rotations $\psi_x \rightarrow e^{\, i \epsilon \gamma_{5,x}} \psi_x, 
\overline{\psi}_x \rightarrow e^{\, i \epsilon \gamma_{5,x}} \overline{\psi}_x$ where $\gamma_{x,5} = (-1)^{x_1+x_2}$. Thus massless modes are not
protected by chiral symmetry such that the quartic term generates a mass, which in turn allows for a non-trivial $\theta$-dependence.

Charge conjugation symmetry acts on the dynamical variables as
\begin{equation}
A_{x,\mu} \; \rightarrow \; - \, A_{x,\mu} \; , \; \;  n_{x} \; \rightarrow \; - \, n_{x} \; , \; \; 
\psi_x \; \rightarrow \; \overline{\psi}_x \; , \; \; \overline{\psi}_x \; \rightarrow \; \psi_x \; ,
\end{equation}
which leaves the fermion action $S_F$ and the quadratic term in the Villain Boltzmann factor invariant, but not the topological term. 
The integer-valued topological charge $Q = \sum_x n_x$ changes sign, such that for $\theta = \pi$ we indeed find
charge conjugation symmetry of the whole system (which of course also holds for the trivial case $\theta = 0$). 

It is obvious that for $\theta \neq 0$ the topological term in (\ref{eq:Bfactor}) introduces a complex action problem. 
However, this complex action problem can be overcome by switching to a worldline/worldsheet representation. This representation solves 
the complex action problem from the topological term \cite{Gattringer:2018dlw,Goschl:2018uma} 
(see \cite{Kloiber:2014dfa,Gattringer:2015baa} for an equivalent 
result for the Wilson action), while the remaining sign problem from the Grassmann variables and the staggered signs is known to be absent for 
massless staggered fermions in 2 dimensions \cite{Gattringer:2015nea,Goschl:2017kml}.  The worldline representation of the partition sum,
\begin{eqnarray}
Z  & = & \left( \frac{1}{\sqrt{ 32 \pi \beta}}\right)^{\!V} \!\! \sum_{ \{l,d,p\}} W_{\beta,\theta} [\, p] \; W_J[d] \; 
\prod_{x} \delta\big( (\nabla l)_x \big) \, \prod_{x,\mu} \delta \big( l_{x,\mu} + \varepsilon_{\mu \nu}[p_x - p_{x-\hat\nu} ] \big) 
\label{eq:Z_DL}
\\
&  & \qquad \qquad \qquad \qquad  \times \, 
\prod_{x} 
\delta\!\left( \! \sum_\mu \Big[ d_{x,\mu} \!+\! d_{x-\hat \mu,\mu} + \frac{1}{2} [ \, |l_{x,\mu}| \!+\! |l_{x-\hat \mu,\mu}| \, ] \Big] - 1 \! \right),
\nonumber
\end{eqnarray}
is a sum over configurations of the fermionic flux variables $l_{x,\mu} \in \{-1,0,1\}$ assigned to the links, the link-based dimer variable 
$d_{x,\mu} \in \{0,1\}$ and the plaquette occupation numbers $p_x \in \mathds{Z}$. 

\begin{figure}[t!] 
\begin{center}
\hspace*{-2.1mm}
\includegraphics[scale=1.1,clip]{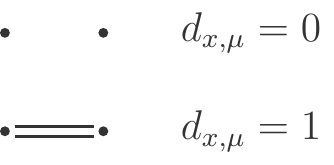} 
\hspace{14mm}
\includegraphics[scale=1.1,clip]{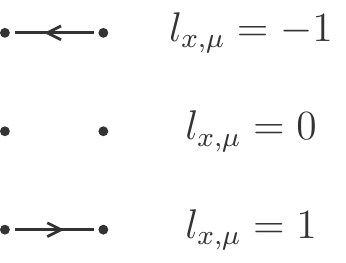} 
\hspace{14mm}
\includegraphics[scale=1.1,clip]{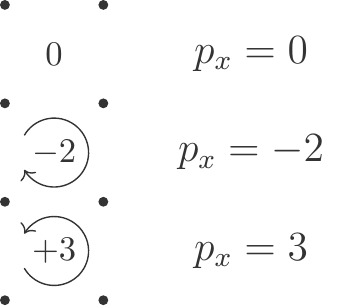} 
\end{center}
\caption{Graphical illustration of the elements of the worldline/worldsheet representation.}
\label{figure_1}
\end{figure}

The dynamical variables in the worldline/worldsheet representation (\ref{eq:Z_DL}) come with constraints that are expressed with Kronecker 
deltas here denoted as $\delta(n) = \delta_{n,0}$. The first constraint implements a zero divergence condition for the fermion fluxes, 
$(\nabla l)_x  = \sum_\mu [l_{x,\mu} - l_{x-\hat \mu, \mu}] = 0 \; \forall x$, which implies that the fermion flux variables must form oriented closed loops.
The second constraint comes from gauge invariance and implies that at each link the combined flux from the fermion loops and the plaquette 
occupation numbers vanishes. Finally, the third constraint is a consequence of integrating out the Grassmann variables and requires that each 
site is either the endpoint of a dimer or is run through by one unit of fermion flux. 

\begin{figure}[t!!] 
\begin{center}
\hspace*{-2.1mm}
\includegraphics[scale=1,clip]{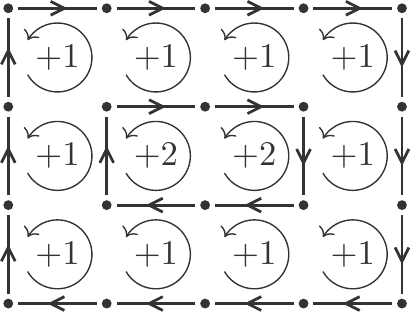} 
\hspace{12mm}
\includegraphics[scale=1,clip]{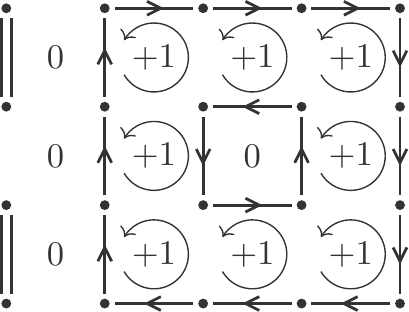} 
\hspace{12mm}
\includegraphics[scale=1,clip]{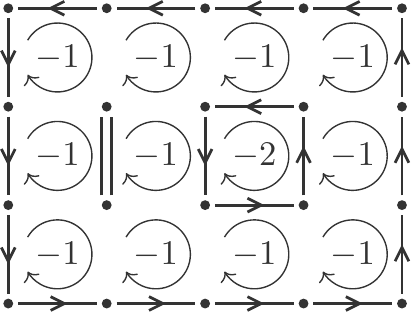} 
\end{center}
\caption{Examples of admissible configurations of dimers (double lines), fermion loops (single lines with arrows) and plaquette occupation numbers 
(oriented circular fluxes with numerical values of plaquette flux).}
\label{figure_2}
\end{figure}

We will use a graphical representation of these
variables where the fermionic flux is represented by oriented lines, the dimers by double lines and the plaquette occupation numbers by 
circular fluxes with the amount of flux written out explicitly (compare Fig.~\ref{figure_1}).
In Fig.~\ref{figure_2} we show examples of admissible configurations of
fermion flux, dimers and plaquette occupation numbers that obey the three constraints. 

In (\ref{eq:Z_DL}) all configurations are assigned weight factors $W_{\beta,\theta} [\, p]$ and $W_J[d]$ given by 
\begin{equation}
W_{\beta,\theta} [\, p] \; = \; \prod_x e^{\, - \, \frac{1}{2\beta} \big(p_x + \frac{\theta}{2\pi} \big)^2 }
\quad , \qquad W_J[d] \; = \; \prod_{x,\mu} (1+ J)^{d_{x,\mu}} \; .
\label{weights}
\end{equation}
Obviously, for all values of the couplings $\beta, \theta$ and $J > -1$ the weight factors are real and positive, such that the complex action problem is
completely solved. In terms of the flux and plaquette occupation numbers the charge conjugation symmetry operates as 
$l_x \rightarrow - l_x$ and $p_x \rightarrow - p_x$ for $\theta = 0$, while for $\theta = \pi$, the plaquette occupation numbers transform as
$p_x \rightarrow - p_x - 1$.  

\section{Simulating the worldline representation}

As was outlined above, we will simulate the system using the worldline/worldsheet representation in order to overcome the complex action problem 
of the standard representation. 
Before we come to the discussion of Monte Carlo update strategies for the highly constrained variables of the worldline/worldsheet representation, 
we also need to express the physical observables in terms of the new degrees of freedom. All observables we consider here are derivatives 
of $\ln Z$ with respect to the couplings. These derivatives can be computed both, in the original and in the worldline/worldsheet representation, where 
the latter gives the expressions in terms of the new variables. More specifically, we here consider the topological charge density $\langle q \rangle$ with
$q = Q/V$,
the corresponding susceptibility $\chi_q$ and the gauge action density $\langle F^2 \rangle$. The corresponding definitions and worldline expressions 
are given by (in the expression for the topological charge density we dropped a factor $i$, such that the second expression in (3.1) is a definition ensuring 
that $\langle q \rangle$ is real)
\begin{eqnarray}
\hspace*{-8mm} && \langle q \rangle \; = \; - \frac{1}{V} \frac{\partial}{\partial \theta} \ln Z \; = \; \frac{1}{2\pi \beta V} \left \langle 
\!\sum_x  \left( p_x \!+\! \frac{\theta}{2\pi} \! \right) \!\right \rangle \; , 
\label{q_dual}
\\
\hspace*{-8mm}  && \chi_{q} \; = \; \frac{\partial}{\partial \theta} \langle q \rangle \; = \; 
\frac{1}{4\pi^2 \beta}  - \frac{1}{4\pi^2 \beta^2 V} \left \langle \left[
\sum_x  \left( p_x \!+\! \frac{\theta}{2\pi} \! \right)  - \left \langle 
\!\sum_x  \left( p_x \!+\! \frac{\theta}{2\pi} \! \right) \! \right \rangle \right]^{\!2} \, \right\rangle  \; ,
\\
\hspace*{-8mm}&& \big\langle F^2 \big\rangle \; = \; - \frac{2}{V} \frac{\partial}{\partial \beta} \ln Z \; = \; \frac{1}{\beta} - \frac{1}{\beta^2 V} \left \langle 
\!\sum_x \! \left( p_x \!+\! \frac{\theta}{2\pi} \! \right)^{\!2} \right \rangle \; . 
\end{eqnarray}

\begin{figure}[t] 
\begin{center}
\includegraphics[scale=0.9,clip]{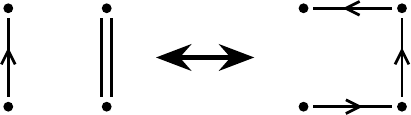} 
\hspace{20mm}
\includegraphics[scale=0.9,clip]{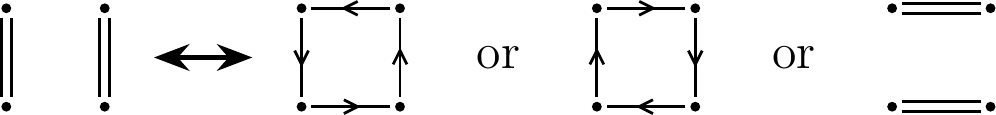}  
\end{center}
\vspace*{-5mm}
\caption{Graphical representation of the local update steps we use. Note that also the plaquette occupation numbers need to be changed when the flux
around the plaquette is altered (not shown here).}
\label{figure_3}
\end{figure}

The main challenge of a simulation with worldlines and worldsheets is to find updates that are ergodic and at the same time keep all constraints intact,
i.e., the fermion fluxes form closed oriented non-intersecting loops, each site is either run through by a fermion loop or is the endpoint of a dimer and finally
on all links the fluxes from the fermion lines and the plaquette occupation numbers compensate each other. We solve this challenge by combining local 
updates and dimer worms \cite{Gattringer:2018dlw,Goschl:2018uma,Gattringer:2015baa,Gattringer:2015nea,Goschl:2017kml}. 
The local updates may deform loops by adding or removing dimers parallel to a loop segment. The local updates may furthermore 
exchange two parallel loops with a fermion loop around the respective plaquette (the orientation of the loop is chosen randomly) or replace the two dimers 
by two perpendicular ones (compare the illustration in Fig.~\ref{figure_3}). Note that in all cases where the fermion flux 
around the plaquette changes, the plaquette occupation number also changes 
by one unit to keep the flux compensation constraint intact. The various changes are accepted with a Metropolis decision, where the acceptance probability 
depends on $J$ when a change of the number of dimers is involved and also on $\beta$ and $\theta$ when the plaquette occupation number has to change. 

\begin{figure}[b] 
\begin{center}
\hspace*{-2mm}
\includegraphics[scale=0.68,clip]{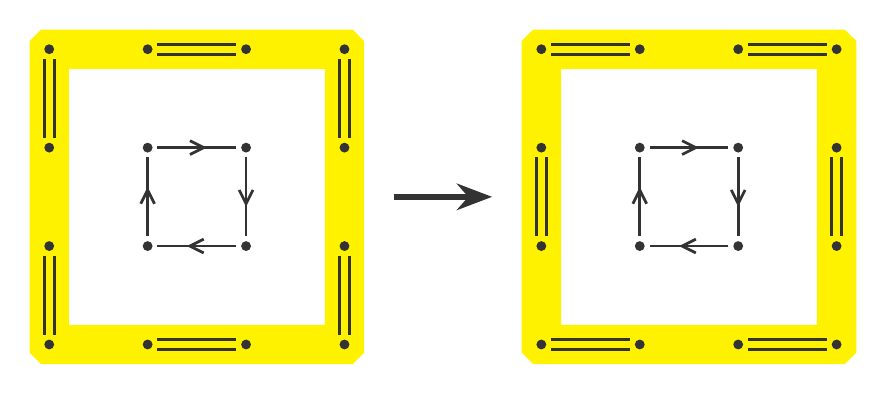} 
\hspace{18mm}
\includegraphics[scale=0.68,clip]{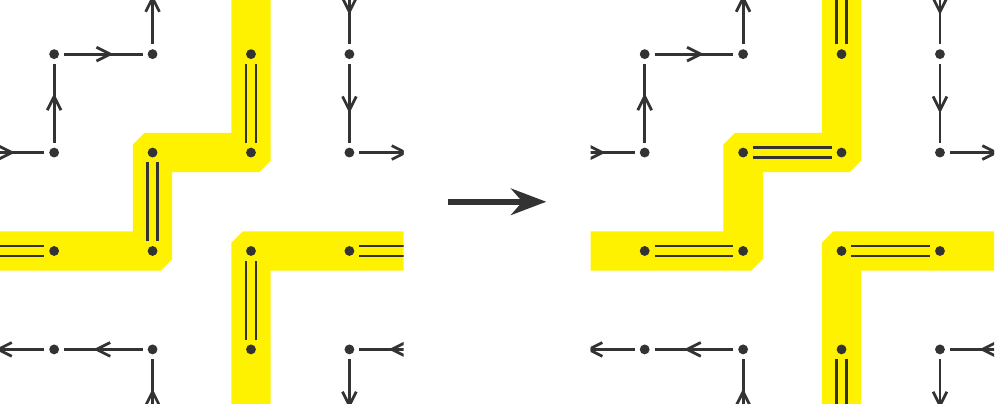}  
\end{center}
\vspace*{-5mm}
\caption{Graphical representation of the dimer worms. The yellow band is the closed worm contour and the two pairs of plots illustrate how the
dimers are shifted along the contour. Note that the worm contour also can wind around the periodic boundary conditions, as is illustrated in the rhs.~plot.}
\label{figure_4}
\end{figure}

The second key element of our update strategy are dimer worms. A worm is started at the endpoint of a dimer and then attempts to propagate along a 
chain of dimers, choosing new directions at random. In case this chain forms a closed contour, the worm terminates and 
all dimers on the contour are shifted by one unit along the contour
(compare Fig.~\ref{figure_4}). Note that the contours may also wind around the compact boundary conditions, as is illustrated in the rhs.\ example in 
Fig.~\ref{figure_4}. We remark, that for the dimer worms the Boltzmann weight remains unchanged and no Metropolis or other weighted decision step is 
necessary. The local plaquette-based updates and the dimer worms already constitute an ergodic 
update that obeys all the constraints.

Finally, we also allow for global updates of all plaquettes by $\pm 1$, which is a step that is not necessary for ergodicity, but helps to faster decorrelate the 
topological charge. This step is again accepted with a Metropolis decision. 

It is obvious, that such a complex algorithm for highly constrained variables needs to be tested thoroughly.  On small lattices one may sum up either all 
or the dominant fermionic configurations (loops and dimers) and then sum over all configurations of the plaquette variables that are compatible with the 
given set of fermion loops. This is an infinite sum, but due to the Gaussian nature of the weight  $W_{\beta,\theta}$  
(compare Eq.~(\ref{weights})), this sum is fast converging and may be truncated after only a few terms. In this way one can generate analytic results for the 
observables that may be used as reference data. Using $4 \times 4$ lattices we computed all first and second derivatives of $\ln Z$ as observables
and evaluated them as a function of all couplings $\beta, \theta$ and $J$. Our simulation results were compared to these reference data and we found 
excellent agreement for all values of the couplings, indicating that the algorithm is implemented correctly.

In our main simulations we work on $L \times L$ lattices with the linear extent $L$ ranging between $L = 8$ and $L = 40$. 
We combine 5 updates of all plaquettes 
with the local update, 10 dimer worms, and 3 global updates of all plaquette occupation 
numbers into one sweep. We typically use $10^9$ sweeps for equilibration followed by $10^7$ to $10^8$ measurements of our observables  
separated by $10^2$ sweeps for decorrelation.
All errors we show are statistical errors determined with the Jackknife method combined with a blocking analysis.

\section{First numerical results}

\begin{figure}[t] 
\begin{center}
\hspace*{-3mm}
\includegraphics[scale=0.75,clip]{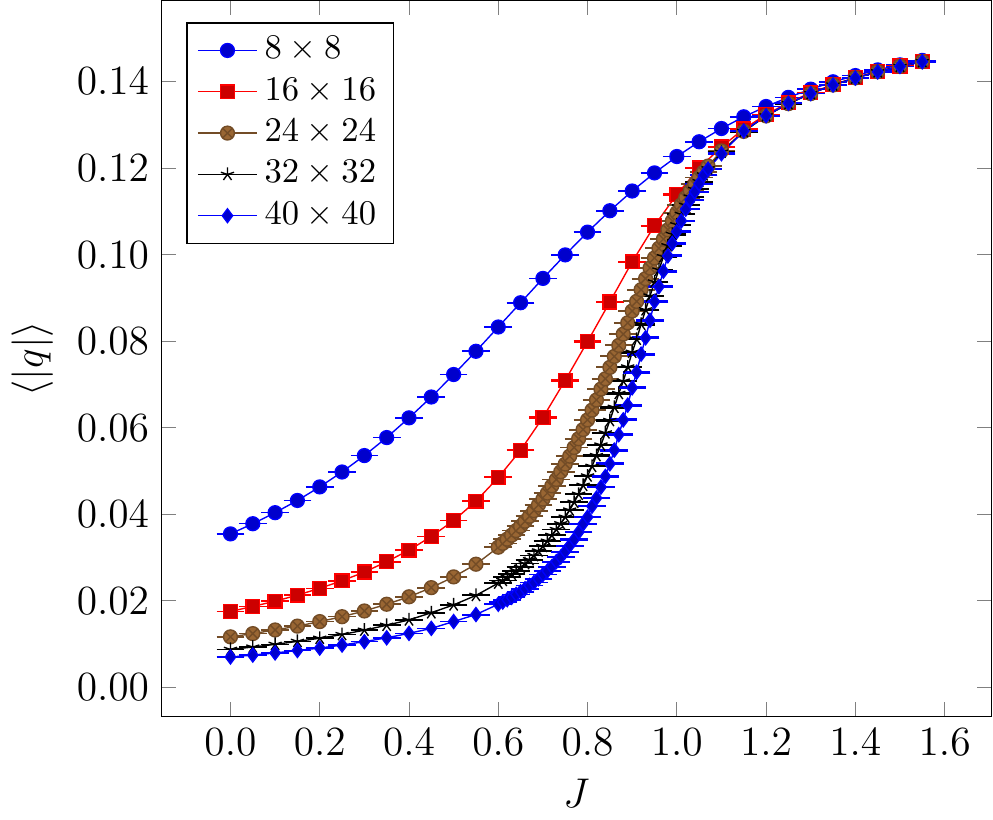} 
\hspace{2mm}
\includegraphics[scale=0.75,clip]{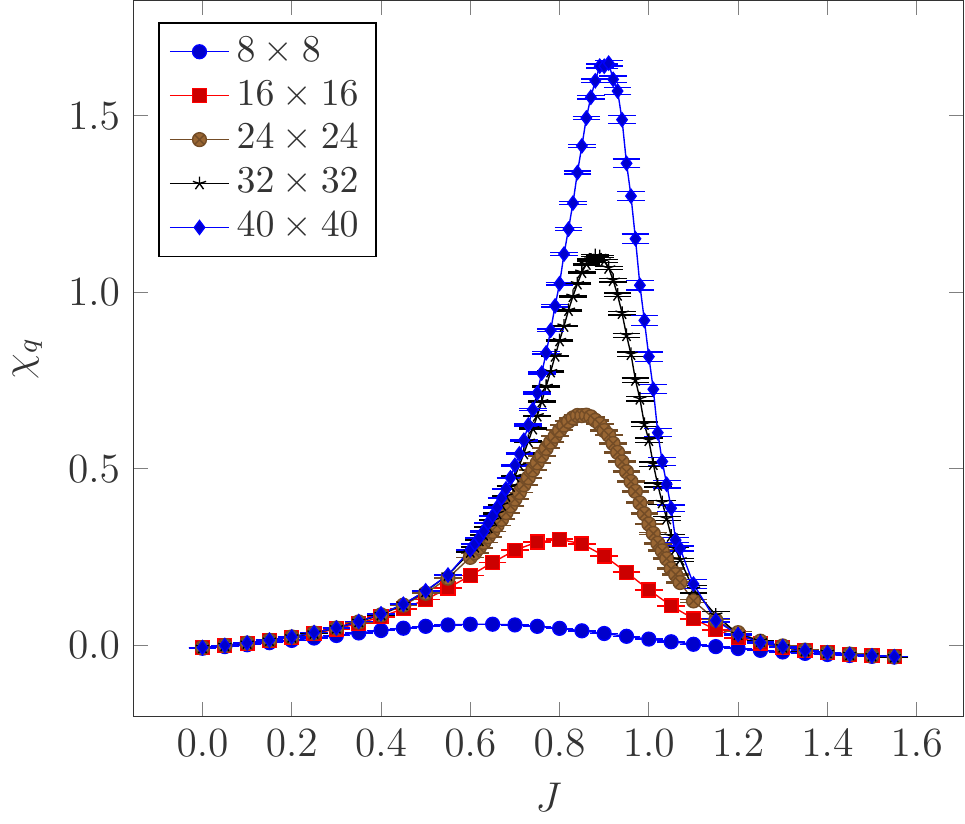}  
\end{center}
\vspace*{-7mm}
\caption{Vacuum expectation value of the absolute value of the topological charge density $\langle | q | \rangle$ (lhs.\ plot) 
and the corresponding susceptibility $\chi_q$ (rhs.). 
Both observables are shown as a function 
of the quartic coupling $J$ at fixed values $\theta = \pi$, $\beta = 0.5$, and we compare results for different lattice sizes.}
\label{figure_5}
\end{figure}

We begin the discussion of our (preliminary) numerical results with the topological charge density $\langle q \rangle$ with $q = Q/V$. Since the
topological charge $Q$ is odd under charge conjugation the charge density is a suitable order parameter to study possible breaking of charge conjugation 
at $\theta = \pi$, i.e., the non-trivial implementation of this symmetry. In the lhs.\ plot of Fig.~\ref{figure_5} we show the expectation value 
$\langle | q | \rangle$, where the absolute value is used to observe signals of symmetry breaking also on a finite volume. The figure shows 
$\langle | q | \rangle$ for $\beta = 0.5$ and $\theta = \pi$ as a function of the quartic coupling $J$, and we compare different volumes. The figure 
shows that for weak coupling the data approach zero in the thermodynamical limit, signaling an unbroken charge conjugation symmetry. At strong
coupling $J$ the values of $\langle | q | \rangle$ converge to a constant value indicating that charge conjugation symmetry is broken there. The largest slope 
for $\langle | q | \rangle$ is observed near $J_c \sim 0.9$, indicating that the transition is located in this region. 

In the rhs.~plot of Fig.~\ref{figure_5} we show the corresponding susceptibility $\chi_q$. It is obvious that $\chi_q$ develops a well pronounced maximum 
with increasing volume which is located at $J_c \sim 0.9$, giving our current estimate for the critical point where charge conjugation symmetry is broken. 
The height of the maximum is scaling with the volume, indicating that we observe a 
true phase transition. We are currently conducting a finite size scaling analysis of $\chi_q$, as well as of the Binder cumulant for the topological charge
to determine the nature of the transition and its critical exponents\footnote{Actually the transition is expected to 
be in the 2d Ising universality class, and a first preliminary analysis of the volume scaling of the maxima of $\chi_q$ indeed shows a 
$L^{7/4}$ behavior as in the Ising case.}.

The dual representation not only admits a Monte Carlo simulation without complex action problem, but also allows for an elegant understanding 
of the physics in terms of the worldline/ worldsheet representation. For $J \rightarrow \infty$ the lattice is completely filled with 
dimers such that due to the absence of loops the gauge field dynamics decouples from the fermions. The dominant admissible gauge field 
configurations are $p_x = 0 \; \forall x$ and $p_x = - 1 \; \forall x$, such that the sum in (\ref{q_dual}) is either $-V/2$ or $+V/2$ leading to 
$\langle | q | \rangle = 1/4 \pi \beta$. On the
other hand, for $J = -1$ dimers are completely absent and fermions
only can generate loops that are non-intersecting. The gauge field weight factors are largest when $p_x = 0$ or $p_x = -1$, which are those
values that are needed for filling the plaquettes inside positively oriented loops with $p_x = -1$ and keeping the plaquettes outside the loops empty 
($p_x = 0$). The values $p_x = -1$ and $p_x = 0$ come with equal probability and contribute $-1/2$ and $+1/2$ to the sum for $q$ (see (\ref{q_dual})),
such that summing these dominant values of 
the plaquette occupation numbers 
give rise to $q = 0$, and thus $\langle | q | \rangle = 0$.  Between these limiting cases we expect a transition where the $\mathds{Z}_2$ 
charge conjugation symmetry is broken spontaneously.

\begin{figure}[t] 
\begin{center}
\hspace*{-2mm}
\includegraphics[scale=0.71,clip]{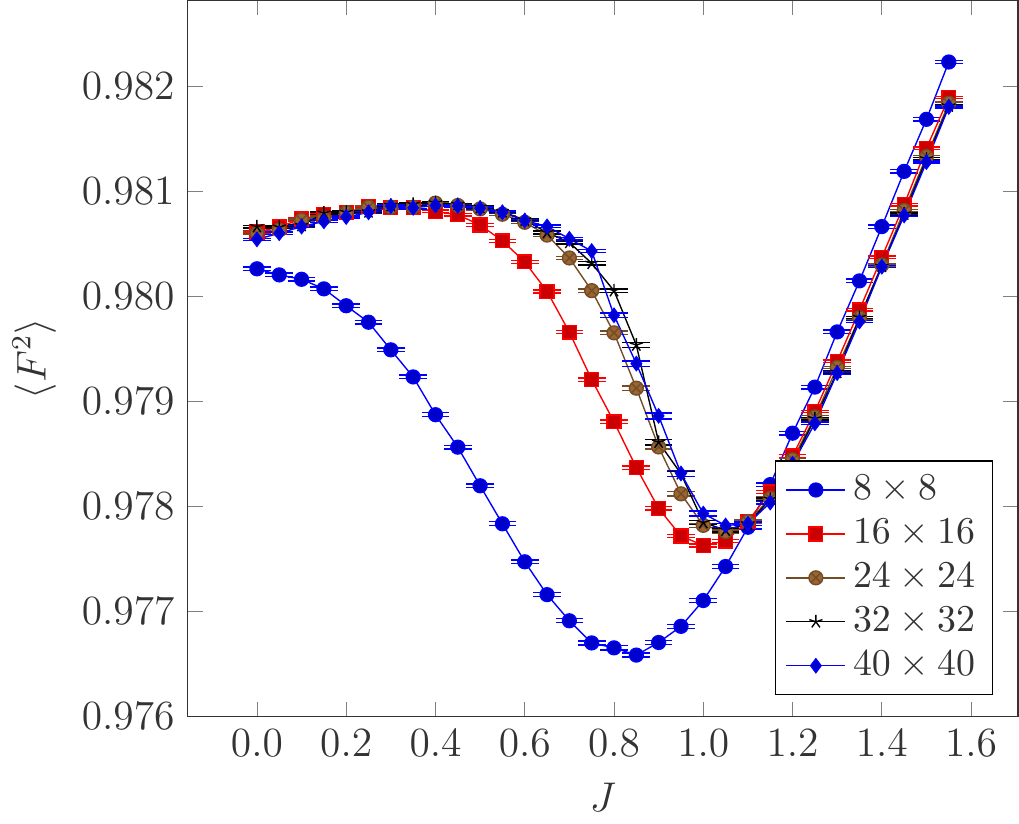} 
\hspace{2mm}
\includegraphics[scale=0.71,clip]{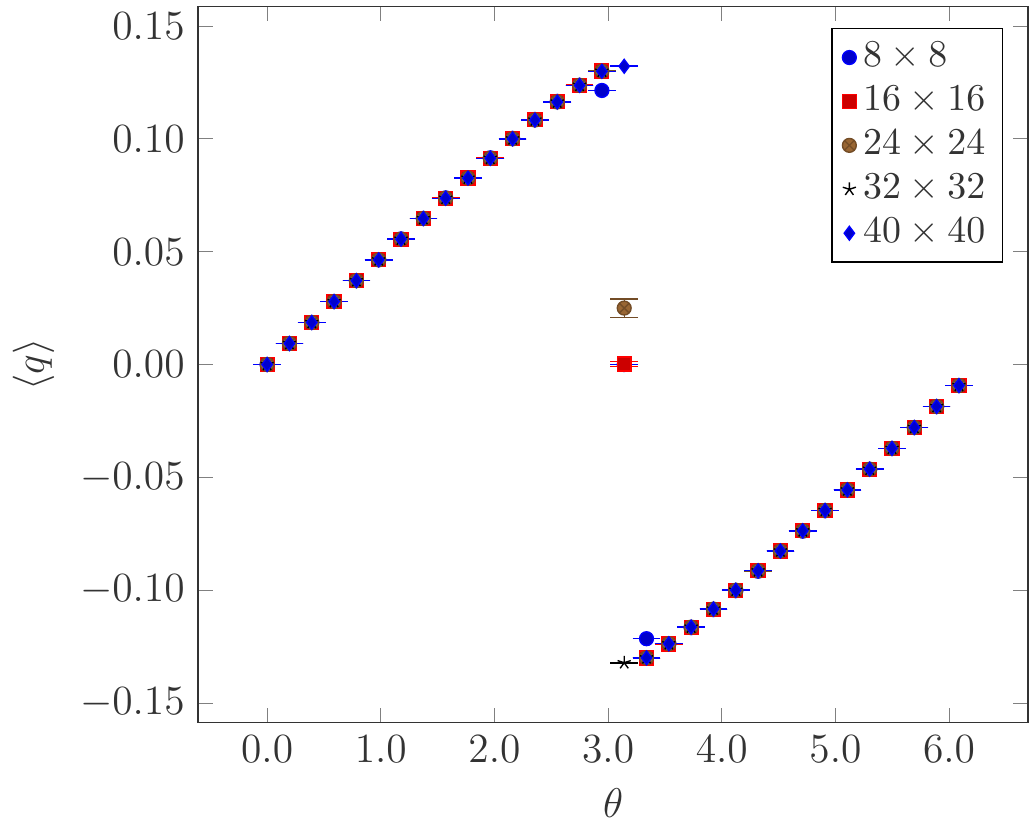}  
\end{center}
\caption{The gauge action density $\langle F^2 \rangle$ as a function of the quartic coupling $J$ (lhs.\ plot) and the topological charge density 
$\langle q \rangle$ as a function of the topological angle $\theta$ (rhs.).}
\label{figure_6}
\end{figure}

To check the consistency of the physical picture, 
in the lhs.~plot of Fig.~\ref{figure_6} we show the expectation value of the gauge field action density $\langle F^2 \rangle$, again for 
$\beta = 0.5$ and $\theta = \pi$ as a function of the coupling $J$. Note that this is an observable that is even under charge conjugation, such that 
we do not expect scaling with the volume. Indeed, we observe only a very small variation of the data in the range $J \in [0.0,1.6]$ (the variation is in 
the per mille range). Furthermore, from lattice size $24 \times 24$ on, the data for different volumes fall on top of each other indicating the 
absence of volume scaling of $\langle F^2 \rangle$.

Finally, in the rhs.~plot of Fig.~\ref{figure_6} we analyze the situation when we allow for explicit breaking of charge conjugation symmetry by setting 
$\theta \neq \pi$. As a matter of fact one may consider the difference $\Delta = \theta - \pi$ as the symmetry breaking parameter, which plays the same 
role as the external magnetic field in the Ising model. The rhs.~plot of Fig.~\ref{figure_6} now shows the expectation value of the topological
charge density $\langle q \rangle$ (note that here the absolute value is absent) as a function of $\theta$, again at gauge coupling  
$\beta = 0.5$. For the quartic
coupling $J$ we use $J = 1.2$, i.e., a value where at $\theta = \pi$ charge conjugation is broken simultaneously. The plot shows that for $\theta = 0$ and $\theta = 2\pi$ we find $\langle q \rangle = 0$, as expected for the points $\theta = 2\pi \, \mathds{Z}$ where charge conjugation symmetry is implemented 
trivially. Near $\theta = \pi$, however, when the symmetry breaking parameter $\Delta = \theta - \pi$ changes sign, 
we find that the order parameter jumps from $\langle q \rangle \sim 0.13$, which is the value 
we also observe at $J = 1.2$ in the lhs.\ plot
of Fig.~\ref{figure_5}, to $\langle q \rangle \sim - \, 0.13$. Obviously we observe
a first order transition when the topological angle $\theta$ crosses $\pi$ in the strong coupling region, i.e., for $J > J_c \sim 0.9$.

Thus we may summarize the phase diagram in the $J$-$\theta$ plane as follows: We find a horizontal first order line at $\theta = \pi$ (or more generally
at $\theta = (2 n + 1) \, \pi$ with $n \in \mathds{Z}$), that starts at $J_c \sim 0.9$ and extends to arbitrarily large quartic coupling $J$. When crossing this 
critical line vertically, i.e., when the symmetry breaking parameter  $\Delta = \theta - \pi$ changes its sign, the order parameter $\langle q \rangle$ 
jumps from positive to negative values. The critical value $J_c \sim 0.9$ appears to be a second order point which is expected to be in the 2d Ising
universality class, a conjecture that we are currently studying numerically.

\vspace{3mm}
\noindent
{\bf Achknowledgments:} We thank Tin Sulejmanpasic for various discussions, in particular about the mechanism for symmetry breaking,
and Uwe Wiese for a discussion concerning the mass generation by the quartic interaction.

\bibliographystyle{utphys}
\bibliography{bibliography}

\end{document}